\documentclass[]{aa}
\usepackage{graphicx}
\usepackage{natbib}
\begin{document}

\title{Effect of a strong magnetic field on the surface electric field of strange stars}

\author{Zheng Xiaoping and Yu Yunwei}

\institute{The Institute of Astrophysics, Huazhong Normal
University, Wuhan 430079, Hubei, China. } \mail{
zhxp@phy.ccnu.edu.cn}
\date{Received..... / Accepted .....}

\abstract{We made a detailed study of the properties of the electron
layer near the quark surface of strange stars with strong ($\sim
10^{14}-10^{17}$G) magnetic fields. The electrostatic potential and
the electric field at the quark surface were calculated as functions
of the magnetic field intensity of bare strange stars. Using an
ultrastrong ($B\geq 2.5\times10^{16}$G) magnetic field, we found
that the distribution of electrons becomes an exponential function
of radial distance, which is quite different in a magnetic
field-free case. We also calculated the variation in gap width
between the strange core and the normal nuclear crust for strange
stars, which is due to magnetic field effect.}

\maketitle

\section*{1. Introduction}
Strange quark matter (SQM), made up of roughly equal numbers of up,
down, and strange quarks, may be the absolute ground state of strong
interaction. If this hypothesis is correct, strange quark stars
(strange star, SS) may exist. Of course, it is very important to
identify SSs. A new method for distinguishing SSs from neutron stars
has been proposed recently based on the strong electric field at the
quark surface of SSs \citep{Xu1,Usov2}.
\par
\citet{Alcock3} first pointed out that a very strong
($\sim10^{17}$V/cm) electric field may exist near the bare quark
surface. Due to the strong electric field, $e^{\pm}$ pairs could be
created spontaneously in a few empty quantum states; these pairs,
which then are thermalized in the electron layer, subsequently
annihilate into photons \citep{Usov4,Usov5}. This mechanism could
play an important role in producing the thermal emission of bare
SSs, because the SQM is a poor radiator of thermal photons at
frequencies less than its plasma frequency
($\sim20$MeV)\citep{Alcock3}. Previous studies
\citep{Alcock3,Kettner8, Huang9} have also shown that the strong
electric field may be decisive for the existence of a nuclear crust
on the quark surface of SSs, which may be created in a supernova or
by accretion from the interstellar medium \citep{Glend6,Glend7}.
This is due to the fact that the strong positive Coulomb barrier
prevents atomic nuclei bound in the nuclear crust from coming into
direct contact with the strange core. A SS with a nuclear crust may
have different cooling time scales and surface properties from a
bare SS. Thus the study of the nuclear crust in a SS is important
for future observations of strange pulsars.
\par
As pulsars, SSs are strongly magnetized. From a sample of more than
400 pulsars, the surface magnetic field intensity lies in the
interval $\sim2\times10^{10}-10^{13}$G  \citep{Manch10}. Several
authors \citep{Bisn11,Duncan12,Thom13} have proposed two different
physical mechanisms that lead to an amplification of some initial
magnetic fields in a collapsing star, fields as strong as $B\sim
10^{14}-10^{16}$G, or even more. Considering the flux throughout the
stars as a simple trapped primordial flux, the internal magnetic
field may go up to $B\sim 10^{18}-10^{20}$G. The energy of a charged
particle changes significantly in the quantum limit if the magnetic
field is equal to or greater than some critical value
$B_{c}=m_{i}^{2}c^{3}/(q_{i}\hbar)$G, where $m_{i}$ and $q_{i}$ are
the mass and charge, respectively. For $u$ and $d$ quarks of current
mass 5 MeV, $B_{c}\sim4.4\times 10^{15}$G, while for $s$ quark of
current mass 150MeV, it is $\sim10^{18}$G. Especially for electrons
of mass 0.5MeV, a strong magnetic field that exceeds the critical
value $B_{c}\sim4.4\times 10^{13}$G forces the electron to behave as
a one-dimensional gas rather than three-dimensional. Therefore, the
quantum-mechanical effect of the magnetic field on the distribution
of electrons near the quark surface of SSs cannot be neglected, just
as the temperature effect leads to a considerable reduction of the
electrostatic potential at the quark surface \citep{Kettner8}.
\citet{Phukon14} performed a perturbative expansion of the
thermodynamics to the lowest order term in $B^{2}/\mu^{2}$ to reveal
the magnetic effect. However, in his work, the details are lost in
the middle magnetic field intensity, and the approach is invalid for
the ultra-strong field case.
\par
In this work we present a detailed investigation of the properties
of the electron layer in the presence of a magnetic field and probe
the magnetic field dependence of the gap width between the crust and
the quark core. In the next section we describe the thermodynamic
properties of bulk SQM in the presence of a strong magnetic field at
zero temperature. In Sect. 3 we extend the Poisson equation near the
quark surface to include the effect of magnetic field, and determine
the electrostatic potential numerically. For an ultra-strong
magnetic field, we simplify the equations and solve them
analytically. In Sect. 4 we present the results of SSs with a thin
crust. In the last section, we give our conclusions.
\section*{2. Bulk SQM in strong magnetic fields}
For a constant magnetic field $B$, the single particle(species $i$)
energy eigenvalue is written as
\begin{equation}
\varepsilon_{i}=\sqrt{p_{i}^{2}+m_{i}^{2}+(2n+s+1)q_{i}B},
\end{equation}
where $n=$0,1,2,..., are the principle quantum numbers for allowed
Laudau levels, $s=\pm1$ refers to spin up(+) and down(-), and
$p_{i}$ and $q_{i}$ are the components of particle momentum along
the field direction and the absolute value of the electric charge,
respectively. Setting $2n+s+1=2\nu$, where $\nu=$0,1,2,..., we can
rewrite the single particle energy eigenvalue in the following form
\begin{equation}
\varepsilon_{i}=\sqrt{p_{i}^{2}+m_{i}^{2}+2\nu q_{i}B}.
\end{equation}
Now it is very easy to show that the $\nu=0$ state is singly
degenerate, while the others with $\nu\neq0$ are doubly degenerate.
To a zero temperature approximation, the thermodynamic potential of
the $i$th species is given by \citep{Chak15}
\begin{equation}
\begin{array}{cc}
\Omega_{i}=-\frac{g_{i}q_{i}B}{4\pi^{2}}\sum\limits_{\nu_{i}=0}^{\nu_{i,m}}b_{\nu_{i}0}\left\{\mu_{i}\sqrt{\mu_{i}^{2}-m_{i}^{2}-2\nu_{i}
q_{i}B}\hspace{1.2cm}\right.\\
\hspace{0.1cm}\left.-(m_{i}^{2}+2\nu_{i}
q_{i}B)\ln\left[\frac{\mu_{i}+\sqrt{\mu_{i}^{2}-m_{i}^{2}-2\nu_{i}
q_{i}B}}{\sqrt{m_{i}^{2}+2\nu_{i} q_{i}B}}\right]\right\},
\end{array}
\end{equation}
where $g_{i}$ and $\mu_{i}$ are the degeneracy  (equaling 6 for
quarks and 2 for electron) and chemical potential of the $i$th
species, $b_{\nu_{i}0}=1-\frac{1}{2}\delta_{\nu_{i}0}$. The upper
limit $\nu_{i,m}$ is the greatest integer not exceeding
$(\mu_{i}^{2}-m_{i}^{2})/2q_{i}B$.
\par
For a SS, the net positive charge of the quarks will be balanced
locally by electrons up to radial distances $r\leq R_{m}$, where
$R_{m}$ is only slightly smaller than the stellar radius $R$. This
charge neutrality condition of bulk matter is expressed as
\begin{equation}
2n_{u}-n_{d}-n_{s}-3n_{e}=0.
\end{equation}
And the baryon number density of the system is given by
\begin{equation}
n_{B}=\frac{1}{3}(n_{d}+n_{u}+n_{s}),
\end{equation}
which is considered as a constant parameter. For bulk SQM, assuming
the condition of $\beta$ equilibrium, we have
\begin{equation}
\mu_{d}=\mu_{s}=\mu,
\end{equation}
\begin{equation}
\mu_{u}=\mu-\mu_{e}.
\end{equation}
\par
To solve the above equations for a given magnetic field and baryon
number density, we need the relationship between $n_{i}$ and
$\mu_{i}$. From the well-known thermodynamic relation, the
expression for the number density of the $i$th species (i=u,d and e)
is given by \citep{Chak15}
\begin{equation}
n_{i}=-\left(\frac{\partial\Omega_{i}}{\partial\mu_{i}}\right)=\frac{g_{i}q_{i}B}{2\pi^{2}}\sum\limits_{\nu_{i}=0}^{\nu_{i,m}}b_{\nu_{i}0}\sqrt{\mu_{i}^{2}-m_{i}^{2}-2\nu_{i}
q_{i}B}.
\end{equation}
To the weak magnetic field limit
$(\mu_{i}^{2}-m_{i}^{2})/2q_{i}B\rightarrow \infty$, taking
$\sum\limits_{\nu_{i}=0}^{\nu_{i,m}}\rightarrow
\int\limits_{0}^{\nu_{i,m}}d\nu_{i}$, we can obtain the classical
expression of the number density
\begin{equation}
n_{i}=\frac{g_{i}(\mu_{i}^{2}-m_{i}^{2})^{3/2}}{6\pi^{2}},
\end{equation}
which is valid for $s$ quark due to its large mass
\begin{equation}
n_{s}=\frac{(\mu_{s}^{2}-m_{s}^{2})^{3/2}}{\pi^{2}}.
\end{equation}
On the other hand, for an ultra-strong field $B$ that satisfies
$(\mu_{i}^{2}-m_{i}^{2})/2q_{i}B<1$, only the state $\nu_{i}=0$ is
allowed. We then can get rid of the sum for Laudau levels in Eq.(8)
to read
\begin{equation}
n_{i}=\frac{g_{i}q_{i}B}{4\pi^{2}}\sqrt{\mu_{i}^{2}-m_{i}^{2}}.
\end{equation}
Now we can solve Eqs.(4-7) to obtain the number densities and the
chemical potential of all species in the interior including the
vicinity of $R_{m}$ ($r\leq R_{m}$) of SSs with various values of
$B$.
\section*{3 Bare quark surface}
\subsection*{3.1 Equations and numerical results}
Beyond $R_{m}$, in the region $R_{m}\leq r\leq \infty$, the charge
neutrality is a global rather than a local condition. A concise
model for this layer was first developed by \citet{Alcock3}. In a
simple Thomas-Fermi model, the electrostatic potential near the
quark surface is described by Poisson's equation
\begin{equation}
\frac{d^{2}V}{dr^{2}}= \left\{\begin{array}{cc} 4\pi
e^{2}(n_{e}-n_{q}),\hspace{0.5cm}r\leq R,\\
4\pi e^{2}n_{e},\hspace{1.7cm}r>R, \end{array}\right.
\end{equation}
where $V/e$ is the electrostatic potential. The chemical equilibrium
of the electrons in the layer implies that the value
$\mu_{\infty}=\mu_{e}-V$ is constant. Since far outside the star,
both $V$ and $\mu_{e}$ tend to 0, it follows that $\mu_{\infty}=0$
and $\mu_{e}=V$, and
\begin{equation}
n_{q}=\frac {2}{3}n_{u}-\frac{1}{3}n_{d}-\frac{1}{3}n_{s}
\end{equation}
 is the quark charge density inside the quark surface. Because $R$ and
$R_{m}$ differ only by a few hundred fermi, density $n_{q}$ in this
range can be treated as being independent of $r$. Because of the
local electric charge neutrality inside $R_{m}$,
\begin{equation}
n_{q}=n_{e}(
R_{m})=\frac{g_{e}eB}{2\pi^{2}}\sum\limits_{\nu=0}^{\nu_{m}}b_{\nu0}\sqrt{V_{q}^{2}-m_{e}^{2}-2\nu
eB},
\end{equation}
where $V_{q}$ is clearly the chemical potential corresponding to the
net charge density of quarks inside $R_{m}$, which can be obtained
from Eqs.(4-7).
\par
Assuming the magnetic field in the electron layer is uniform and
substituting Eq.(8) into Eq.(12), we come to
\begin{equation}
\frac{d^{2}V}{dr^{2}}=
 \left\{\begin{array}{cc}
\frac{2g_{e}e^{3}B}{\pi}\left(\sum\limits_{\nu=0}^{\nu_{m}}b_{\nu0}\sqrt{V^{2}-m_{e}^{2}-2\nu
eB}\right.\hspace{1.5cm}\\
\hspace{0.5cm}\left.-\sum\limits_{\nu^{'}=0}^{\nu_{m}^{'}}b_{\nu^{'}0}\sqrt{V_{q}^{2}-m_{e}^{2}-2\nu^{'}
eB}\right),\hspace{0.2cm}r\leq R,\\
\frac{2g_{e}e^{3}B}{\pi}\sum\limits_{\nu=0}^{\nu_{m}}b_{\nu0}\sqrt{V^{2}-m_{e}^{2}-2\nu
eB},\hspace{0.7cm}r>R.
\end{array}\right.
\end{equation}
The boundary conditions for the above equation are $V\rightarrow
V_{q}$ and $dV/dr\rightarrow0$ as $r\rightarrow 0$, and
$V\rightarrow 0$ and $dV/dr\rightarrow0$ as $r\rightarrow \infty$,
respectively. By integrating Eq.(15) we can obtain
\begin{equation}
\frac{dV}{dr}=
 \left\{\begin{array}{cc}
 -\sqrt{\frac{2g_{e}e^{3}B}{\pi}}\left\{\sum\limits_{\nu=0}^{\nu_{m}}b_{\nu0}\left [V\sqrt{V^{2}-m_{e}^{2}-2\nu
eB}\right.\right.\\
-V_{q}\sqrt{V_{q}^{2}-m_{e}^{2}-2\nu eB}\\
\left.-(m_{e}^{2}+2\nu
eB)\ln\left(\frac{V+\sqrt{V^{2}-m_{e}^{2}-2\nu
eB}}{V_{q}+\sqrt{V_{q}^{2}-m_{e}^{2}-2\nu eB}}\right)\right ]\\
\left.-2\sum\limits_{\nu^{'}=0}^{\nu_{m}^{'}}b_{\nu^{'}0}(V-V_{q})\sqrt{V_{q}^{2}-m_{e}^{2}-2\nu^{'}
eB}\right\}^{\frac{1}{2}},\\
\hspace{5cm}r\leq R,\\
-\sqrt{\frac{2g_{e}e^{3}B}{\pi}}\left\{\sum\limits_{\nu=0}^{\nu_{m}}b_{\nu0}\left
[V\sqrt{V^{2}-m_{e}^{2}-2\nu eB}\right.\right.\\
\left.\left.-(m_{e}^{2}+2\nu
eB)\ln\left(\frac{V+\sqrt{V^{2}-m_{e}^{2}-2\nu
eB}}{\sqrt{m_{e}^{2}+2\nu eB}}\right)\right
]\right\}^{\frac{1}{2}},\\
\hspace{5cm}r>R.

\end{array}\right.
\end{equation}
\par
The global charge neutrality determines the potential at the quark
surface to be \citep{Kettner8}
\begin{equation}
V(R)=V_{q}-\frac{P_{e}(R_{m})-P_{e}(\infty)}{n_{q}},
\end{equation}
\begin{figure}
\resizebox{\hsize}{!}{\includegraphics{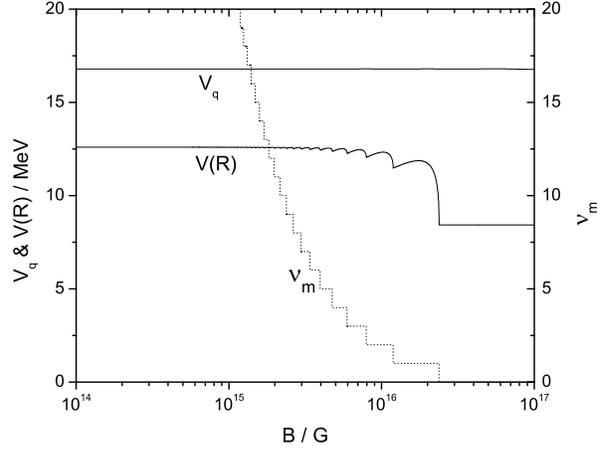}}
\caption{$V_{q}$ and $V(R)$ vs. magnetic field intensity $B$ for
$n_{B}=2.5n_{0}$. The dotted curve shows the number of the maximum
Laudau level $\nu_{m}$ with different $B$.}
\end{figure}
\begin{figure}
\resizebox{\hsize}{!}{\includegraphics{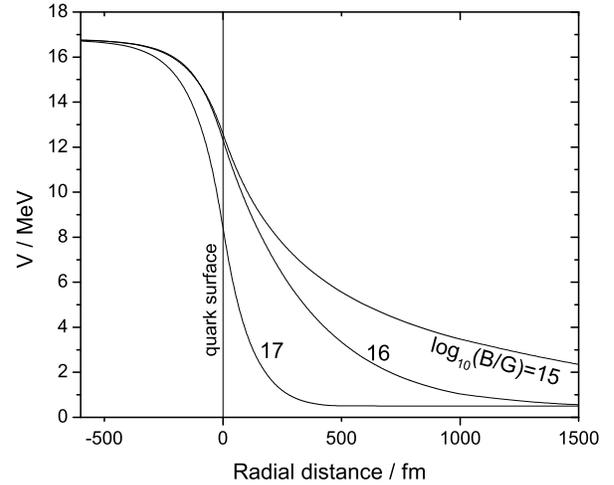}}
\caption{Electrostatic potential of electrons in the close
vicinity inside and outside of the quark surface of Bare SSs. The
location of the surface is indicated by the vertical line. The
figures assigned to these curves refer to magnetic field
intensity.}
\end{figure}
\begin{figure}
\resizebox{\hsize}{!}{\includegraphics{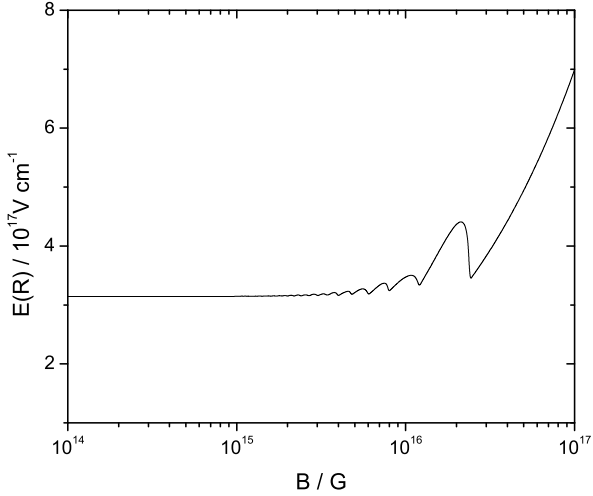}}
\caption{Electric field at the quark surface vs. magnetic field
intensity $B$.}
\end{figure}
where $P_{e}(R)$ is the kinetic pressure of electrons, which could
be obtained by the thermodynamic relation $P_{e}=-\Omega_{e}$. In
Fig.1 we show the variation of $V_{q}$ and $V(R)$ with the magnetic
field intensity $B$. We can see $V_{q}$ is nearly independent of $B$
for our range of fields ($\sim10^{14}-10^{17}$G). And the relation
$V(R)\simeq\frac{3}{4}V_{q}$ applied widely in previous studies is a
good approximation for a weak ($B<10^{15}$G) magnetic field, while
$V(R)\simeq\frac{1}{2}V_{q}$ for an ultra-strong
($B>2.5\times10^{16}$G) field. In the middle range
($B\sim10^{15}-2.5\times10^{16}$G), $V(R)$ oscillates with magnetic
field intensity. These oscillations are easily understood in
accordance with Laudau levels. As is known, the maximum level
$\nu_{m}$, the integer of the ratio
$\frac{(\mu_{e}-m_{e})^{2}}{2eB}$, takes separate values as a
function of $B$. We also show the variation of $\nu_{m}$ with $B$ in
Fig.1. It is obvious that the oscillation of $V(R)$ occurs
synchronously following the skip of $\nu_{m}$. For a weak field,
$\nu_{m}$ has a very large value as
$\frac{\Delta\nu}{\nu_{m}}\rightarrow 0$ so that the oscillation
amplitude is very small. However, the quantum effect becomes
important due to the low value of $\nu_{m}$ that has the comparable
magnitude of $\Delta\nu$ in the middle field intensity. Therefore
$V(R)$ oscillates considerably in this range depending on the
separate values of $\nu_{m}$. The oscillation disappears until
$\nu_{m}$ vanishes for very strong magnetic fields. Furthermore, not
only $V(R)$, but also other quantities involving the summation of
Laudau levels show oscillations (see subsequent figures).
\par
The behaviors of $V(r)$ that vary with respect to $B$ are exhibited
in Fig.2. We can see the stronger the magnetic field is, the larger
the potential changes with $r$. To illustrate the details, we plot
the curve of the electric field (at the quark surface
$E(R)=(-dV/dr)_{r=R}$) versus magnetic field in Fig.3. With the
increase in $B$, the electric field increases remarkably after
oscillations within a range ($\sim10^{15}-2.5\times10^{16}$G).
\subsection*{3.2 Ultra-strong magnetic field approximation} At
ultra-strong magnetic field ($B\geq 2.5\times10^{16}$G for
$V_{q}\sim 17$MeV), only the lowest Laudau level is allowed, and
Eq.(16) becomes to
\begin{equation}
\frac{dV}{dr}=
 \left\{\begin{array}{cc}
 -\sqrt{\frac{2e^{3}B}{\pi}}\times\hspace{4cm}\\
 \left [V\sqrt{V^{2}-m_{e}^{2}}-(2V-V_{q})\sqrt{V_{q}^{2}-m_{e}^{2}}\right.\\
\left.-m_{e}^{2}\ln\left(\frac{V+\sqrt{V^{2}-m_{e}^{2}}}{V_{q}+\sqrt{V_{q}^{2}-m_{e}^{2}}}\right)
\right]^{\frac{1}{2}},\hspace{0.7cm}r\leq R,\\
-\sqrt{\frac{2e^{3}B}{\pi}}\times\hspace{4cm}\\
\left
[V\sqrt{V^{2}-m_{e}^{2}}-m_{e}^{2}\ln\left(\frac{V+\sqrt{V^{2}-m_{e}^{2}}}{m_{e}}\right)\right
]^{\frac{1}{2}}.\\
\hspace{5cm}r>R.
\end{array}\right.
\end{equation}
Since $V\gg m_{e}$, the above equation can be written more simply,
if we ignore the mass of the electron, to read
\begin{equation}
\frac{dV}{dr}=
 \left\{\begin{array}{cc}
 -\sqrt{\frac{2e^{3}B}{\pi}}(V_{q}-V),\hspace{1cm}r\leq R,\\
-\sqrt{\frac{2e^{3}B}{\pi}}V,\hspace{1cm}r>R.
\end{array}\right.
\end{equation}
By integrating this equation and using the relation $V(R)\simeq
\frac{1}{2}V_{q}$ (obtained from Eq.(17) for ultra-strong field
approximation), we get the expression of $V$, and $E=-dV/dr$ as a
function of radial distance $r$
\begin{equation}
V(r)=
 \left\{\begin{array}{cc}
V_{q}\left\{1-\frac{1}{2}\exp\left[\sqrt{\frac{2e^{3}B}{\pi}}(r-R)\right]\right\},\hspace{0.2cm}r\leq R,\\
\frac{1}{2}V_{q}\exp\left[-\sqrt{\frac{2e^{3}B}{\pi}}(r-R)\right],\hspace{0.2cm}r>R.\\
\end{array}\right.
\end{equation}
\begin{equation}
E(r)=
 \left\{\begin{array}{cc}
\frac{1}{2}\sqrt{\frac{2e^{3}B}{\pi}}V_{q}\exp\left[\sqrt{\frac{2e^{3}B}{\pi}}(r-R)\right],\hspace{0.2cm}r\leq R,\\
\frac{1}{2}\sqrt{\frac{2e^{3}B}{\pi}}V_{q}\exp\left[-\sqrt{\frac{2e^{3}B}{\pi}}(r-R)\right],\hspace{0.2cm}r>R.\\
\end{array}\right.
\end{equation}
Substituting Eq.(20) into Eq.(11) in a zero electron mass
approximation, we get the number density of the electron
\begin{equation}
n_{e}(r)=
 \left\{\begin{array}{cc}
\frac{eB}{2\pi^{2}}V_{q}\left\{1-\frac{1}{2}\exp\left[\sqrt{\frac{2e^{3}B}{\pi}}(r-R)\right]\right\},r\leq R,\\
\frac{eB}{4\pi^{2}}V_{q}\exp\left[-\sqrt{\frac{2e^{3}B}{\pi}}(r-R)\right],\hspace{0.2cm}r>R.\\
\end{array}\right.
\end{equation}
Equations (20-22) show that the electrostatic potential, electric
field, and electron number density are all exponential functions of
radial distance. They are quite different from the classical
power-law expressions \citep{Kettner8}.
\section*{4 Thin crust}
The large outwardly directed electric field is capable of supporting
some normal material, which then gives birth to a thin crust. Two
regions should be distinguished just outside the quark surface. The
first one extends from the surface at $r=R$ to the radial distance
where the inner nuclear crust (the crust's base) begins, denoted
$r=R_{c}$. The associated width, $\Delta R\equiv R_{c}-R$, is
referred to henceforth as the gap. The second region begins at
$R_{c}$ and extends in the radial outward direction toward infinity.
Then the corresponding Poisson equation reads
\begin{equation}
\frac{d^{2}V}{dr^{2}}=
 \left\{\begin{array}{cc}
4\pi e^{2}(n_{e}-n_{q}),\hspace{0.5cm}r\leq R,\\4\pi
e^{2}n_{e},\hspace{0.8cm}R<r<R_{c}, \\
4\pi e^{2}(n_{e}-n_{ion}),\hspace{0.5cm}R_{c}\leq r,
\end{array}\right.
\end{equation}
where $n_{ion}$ is the positive charge density of ions. For
simplicity and convenience, we assume the electrostatic potential in
the nuclear crust regime ($r\geq R_{c}$) is constant, as several
authors \citep{Alcock3,Kettner8} have done, i.e., $V(r\geq
R_{c})\equiv V_{c}(=const)$. Then the boundary conditions become
$V\rightarrow V_{q}$ as $r\rightarrow R_{m}$ and $V\rightarrow
V_{c}$ as $r\rightarrow R_{c}$.
\par
Since the mass of ions in the crust is too large for our range of
magnetic field, the lattice and positive charge density of ions
should not be influenced by magnetic field. For a crust, we define a
symbolic electrostatic potential $V_{c}^{*}$ as
\begin{equation}
\begin{array}{cc}
\frac{1}{3\pi^{2}}(V_{c}^{*2}-m_{e}^{2})^{3/2}\equiv
\frac{g_{e}eB}{2\pi^{2}}\sum\limits_{\nu=0}^{\nu_{m}}b_{\nu0}\sqrt{V_{c}^{2}-m_{e}^{2}-2\nu
eB}.
\end{array}
\end{equation}
This new symbol denotes the effective potential at the base of the
crust in the absence of a magnetic field. We use it to designate
different crusts.
\par
The numerical results of $V_{q}$, $V(R)$, and $V_{c}$ (for
$V_{c}^{*}=10$MeV) are shown in Fig.4. The conventional equation
$V(R)=\frac{3}{4}V_{q}+\frac{1}{4}\frac{V_{c}^{4}}{V_{q}^{3}}$ is
also correct for weak magnetic fields ($B<10^{15}$G), while
$V(R)=\frac{1}{2}V_{q}+\frac{1}{2}\frac{V_{c}^{2}}{V_{q}}$ for
ultra-strong magnetic field approximation. Using this numerical
result, we show the results of calculations of the electrostatic
potential with respect to different magnetic fields for
$V_{c}^{*}=10$MeV in Fig.5. In previous studies
\citep{Alcock3,Kettner8,Huang9}, a minimum value of $\sim 200$fm is
established as the lower bound on gap width necessary to guarantee
the crust against strong interaction with the star's strange core.
Figure 6 shows the curves of gap width versus magnetic field
intensity for several crusts. If we adopt 200fm as a criterion for
gap width, then the maximum value of $V_{c}^{*}$ may be 9MeV: the
corresponding crust could be stable for all magnetic fields in our
range ($B<10^{17}$G). Our results demonstrate that the gap width of
various crusts converges to a narrow region ($\sim$200-400fm) with
the increase in $B$. For lighter crusts, the width decreases
considerably, i.e. from $\Delta R=1556$fm at $B=10^{15}$G to $\Delta
R=387$fm at $B=10^{17}$G for $V_{c}^{*}=3$MeV.
\par
However, a realistic gap width should be influenced by the
mechanical balance condition \citep{Huang9}. This condition requires
the electron layer above the quark surface to partially penetrates
the crust. The gap width can be over 200fm for only the very lighter
crusts when the penetrated part is deducted from our present widths.
The details will be studied in our further work.
\begin{figure}
\resizebox{\hsize}{!}{\includegraphics{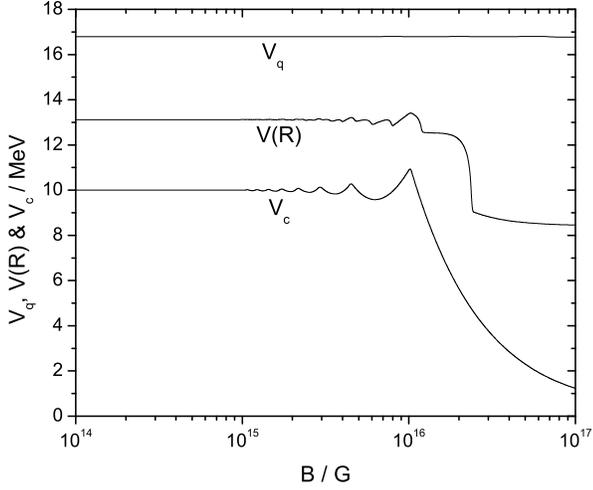}}
\caption{$V_{q}$, $V(R)$ and $V_{c}$ vs. the magnetic field
intensity $B$. A representative value for symbolic electrostatic
potential $V_{c}^{*}=10$MeV.}
\end{figure}
\begin{figure}
\resizebox{\hsize}{!}{\includegraphics{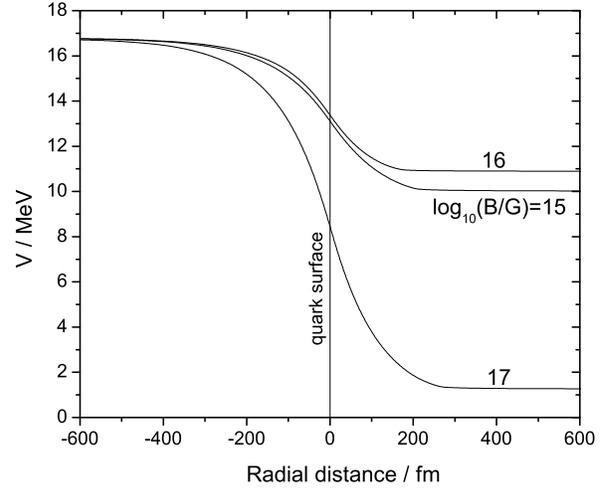}}
 \caption{Same
as Fig.2, but for SSs with a crust with $V_{c}^{*}=10$MeV.}
\end{figure}
\begin{figure}
\resizebox{\hsize}{!}{\includegraphics{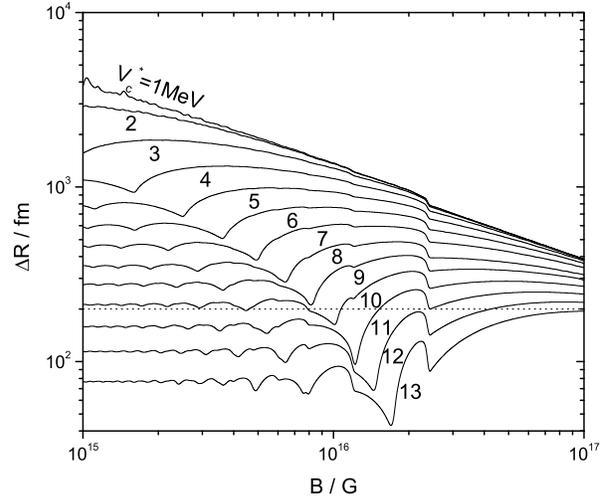}}
 \caption{Gap
width vs. magnetic field intensity. The labels refer to symbolic
electrostatic potential (in MeV).}
\end{figure}
\section*{5.Conclusions}
In this paper, the properties of the electric field near the surface
of strange stars with magnetic field are studied thoroughly. We find
that a weak ($B\leq 10^{15}$G) magnetic field has hardly any effect
on the surface electric field of bare SSs. Secondly, in the middle
range ($B\sim10^{15}-2.5\times10^{16}$G) of magnetic field
intensity, both the electrostatic potential and electric field have
an oscillation as $B$ changes. Thirdly, for ultra-strong
($B\geq2.5\times10^{16}$G) magnetic fields, the electrostatic
potential at quark surface reduces considerably, accompanied by an
increase in the electric field. The electrostatic potential, the
electric field, and the electron number density are all exponential
functions of radial distance, which is quite different from the
field-free case.
\par
We took the effect of magnetic field on the nuclear crust into
account. With the increase in magnetic field intensity, the gap
width of various crusts converges to a narrow region ($\sim
200-400$fm). The lighter crusts have a quick decrease in gap width,
while the heavier ones suffer a slight increase in gap as $B$ is
strengthened. We infer that those may influence the mass of the
crust.
\par
\textit{Acknowledgements.} We would like to thank Dr. Y. F. Huang
for the useful discussion and to acknowledge the support by NFSC
under Grant Nos 10373007 ,90303007, and the Ministry of Education of
China with project No. 704035.

\end{document}